\documentclass[sigconf]{acmart}
\usepackage[utf8]{inputenc}
\usepackage[T1]{fontenc}
\usepackage{lineno}
\usepackage{subcaption}
\usepackage{hyperref}
\usepackage{graphicx}
\usepackage{enumitem}
\usepackage{natbib}
\usepackage{amsfonts}
\usepackage{amsthm}
\usepackage{colortbl}
 \usepackage{booktabs} 
\renewcommand\footnotetextcopyrightpermission[1]{}
\hypersetup{pdfinfo={}}

\definecolor{CorRBelow0}{HTML}{FFFFFF}   
\definecolor{CorR0to24}{HTML}{FFFFFF}    
\definecolor{CorR25to49}{HTML}{FFFFFF}   
\definecolor{CorR50to69}{HTML}{FFFFFF}   
\definecolor{CorR70to89}{HTML}{FFFFFF}   
\definecolor{CorR90to100}{HTML}{FFFFFF}  

\newcommand{\CorrCell}[1]{%
  \begingroup
    \ifdim #1 pt < 0pt
      {#1}%
    \else
      \ifdim #1 pt < 0.25pt
        {#1}%
      \else
        \ifdim #1 pt < 0.50pt
          {#1}%
        \else
          \ifdim #1 pt < 0.70pt
           {#1}%
          \else
            \ifdim #1 pt < 0.90pt
              \it{#1}%
            \else
              \bf{#1}%
            \fi
          \fi
        \fi
      \fi
    \fi
  \endgroup
}

\title{Exploring Economic Sectoral Dynamics Through High-resolution Mobility Data}
\author{Timothy F Leslie}
\orcid{}
\email{tleslie@gmu.edu}
\affiliation{%
    \institution{George Mason University}
    \city{Fairfax}
    \country{USA}
}

\author{Hossein Amiri}
\orcid{0000-0003-0926-7679}
\email{hossein.amiri@emory.edu}
\affiliation{%
    \institution{Emory University}
    \city{Atlanta}
    \country{USA}
}
\author{Andreas Z{\"u}fle}
\orcid{0000-0001-7001-4123}
\email{azufle@emory.edu}
\affiliation{%
    \institution{Emory University}
    \city{Atlanta}
    \country{USA}
}

\begin{document}
\begin{abstract}
    We present a comprehensive dataset capturing patterns of human mobility across the United States from January 2019 to January 2023, based on anonymized mobile device data. Aggregated weekly, the dataset reports visits, travel distances, and time spent at public locations organized by economic sector for approximately 12 million Points of Interest (POIs). This resource enables the study of how mobility and economic activity changed over time, particularly during major events such as the COVID-19 pandemic. By disaggregating patterns across different types of businesses, it provides valuable insights for researchers in economics, urban studies, and public health. To protect privacy, all data have been aggregated and anonymized. This dataset offers an opportunity to explore the dynamics of human behavior across sectors over an extended time period, supporting studies of mobility, resilience, and recovery.

\end{abstract}
\keywords{Human Mobility, Economic Sectors, NAICS CODE}
\maketitle
\pagestyle{plain}

\section*{Background \& Summary}
Scholars have long been intrigued by the intricate dynamics of spatial interaction within communities, especially with regard to how economic activities shape these dynamics \cite{richardson1969regional, delamater2016spatiotemporal, redding2017quantitative, kim2019simulating}. However, a significant challenge in studying these phenomena lies in obtaining accurate and comprehensive data for calibration and theory testing, particularly at the individual level \cite{hurley2024spatiotemporal,amiri2023massive,amiri2024geolife+,amiri2024urban}. A common approach is to utilize cell phone tracking data that has become more accessible and detailed through a variety of data sharing agreements \cite{pesavento2020data}. This shift toward passively collected large-scale data has proven particularly invaluable in understanding human mobility during global disruptions like the COVID-19 pandemic. Such granular data broadens the scope of mobility scholarship and also informs policy decisions, urban planning, and public health strategies by considering how a much larger and more diverse set of individuals responds to economic or environmental shifts\cite{kupfer2021using}.

The COVID-19 pandemic provided an unprecedented opportunity for scholars to explore human mobility patterns amid  widespread disruption. During the early stages of the COVID-19 pandemic, the data company SafeGraph~\cite{safegraph} played a pivotal role by providing complimentary access to high-quality foot-traffic data through its Data for Academics initiative. These data captured visits of individuals (tracked through their mobile devices), their home locations, and their interactions with points of interest such as restaurants or shops. However, SafeGraph later discontinued public access to these data, limiting researchers’ ability to conduct longitudinal analyses. A similar dataset was subsequently published by Advan Research~\cite{advanresearch} and shared through Dewey Data~\cite{deweydata}, offering a valuable continuation of large-scale mobility tracking. Although Advan’s aggregation methods differ slightly, the two datasets share many structural similarities. Our use of Advan data enabled us to extend analysis across the full pandemic and post-pandemic period, capturing trends across a broader set of economic sectors while maintaining consistent measurement of foot-traffic patterns. Recent releases of mobility data have enabled many new data-driven research directions toward understanding change in mobility during a pandemic \cite{chang2021mobility, elarde2021change}. 


Existing studies have demonstrated the utility of aggregated mobility data for measuring origin-to-destination population flows and understanding how people move across different geographic scales \cite{kang2020multiscale}. Building on that foundation, this article takes a sector-focused perspective by aggregating foot-traffic according to the North American Industry Classification System (NAICS). Specifically, for each week and NAICS class of POIs at both a broader (2-digit) and disaggregate (4-digit) levels, our dataset reports the number of individuals visiting a business, the distance they traveled, and the average length of stay. This sector-based approach provides deeper insight into how different types of business establishments experienced changes in visitation over time, offering a valuable complement to origin-destination-based mobility studies. Beyond providing these data for the broader scientific community, we also present an analysis of the data, highlighting patterns in sector-specific behavior and underscoring the importance of utilizing disaggregated sectoral data for many types of research.

For our research, we utilized data from Dewey's Advan Weekly Foot Traffic, spanning from January 7, 2019, to January 2, 2023. This time frame includes periods prior to, during, and following the COVID-19 pandemic, enabling analysis of mobility patterns and behaviors over time.

Organized by NAICS code, this dataset provides detailed information on each POI visited, including number of visitors, the duration of their stay, and the distance of the POIs from their home location. Our dataset includes approximately 7 million of the 12,329,726 total locations. This data is captured for public locations that exclude private residences or other venues not accessible to the general public, allowing us to capture a comprehensive view of consumer- and community-oriented mobility while maintaining analytical consistency across different types of establishments.

\section*{Methods}
Advan foot-traffic data determines each device’s home census block group (CBG) by analyzing its nighttime location over a six-week period using proprietary clustering and location-based algorithms. This approach ensures that the home location of each device is accurately pinpointed, which is crucial for analyzing mobility patterns. To safeguard individual privacy, detail from census block groups where fewer than two devices have visited a specific establishment within a week were excluded. This privacy measure ensures that sensitive location information is not divulged when the number of visitors is too low to maintain anonymity.

For each Point of Interest (POI), we gathered three metrics:
\begin{enumerate}
    \item Total number of visitors: This metric provides a count of unique devices that visited each POI during a week.
    \item Median dwell time: The median duration each visitor spent at the POI, computed from the first detection timestamp to the last detection timestamp at the POI.
    \item Median traveled distance: The median distance of visitors from the visitors' home census block group to the POI, allowing us to understand mobility ranges.
\end{enumerate}
This granular level of detail allows us to understand not only the volume of foot traffic at different locations but also the duration of visits and the distances individuals are willing to travel from their homes. Furthermore, we utilized the category of the POI based on their respective NAICS (North American Industrial Classification System) codes, both at the two digit (broad) and four-digit (more detailed) level. By disaggregating the data in this sectoral manner, we can analyze mobility patterns across various industries, providing insights into how different sectors were affected by the pandemic and subsequent lockdown measures. 

We performed these calculations on a weekly basis throughout the dataset, where each week began on Monday and ended on Sunday. The dataset spans from January 7, 2019, to January 2, 2023, covering both pre-pandemic, pandemic, and post-pandemic periods. This temporal granularity enables us to track trends and observe changes in mobility patterns over time. Weekly aggregation helps in identifying short-term and long-term impacts of events such as the COVID-19 pandemic on human mobility.

\subsection*{Raw Data}
To measure the weekly average distance traveled, the weekly average dwell time, and the weekly average number of visitors, we used foot traffic data from the Weekly Visit Patterns Dataset, provided by Advan Research Corporation \cite{advanresearch} and shared through Dewey Data platform \cite{deweydata} with a paid subscription.


This dataset captures mobility information from approximately 1,000 mobile phone applications spanning various sectors, including social media networks, online retail, healthcare, and navigation. The data captures individual foot-traffic (visited Place of Interest) for about 35 million individual cellphones in the United States, capturing 8 trillion cell data points. The Weekly Visit Patterns Dataset used in this study includes: 
\begin{enumerate}
    \item A list of Points of Interest (POIs) including location and NAICS designation.
    \item For each POI and each week between 01/07/2019 and 01/02/2023: The number of visitors (to the respective POI in the respective week) from each Census Block Group (CBG) (a CBG is a statistical division used in the United States defined to contain between 600 and 3000 individuals), and 
    \item Time and duration of each such visit.
\end{enumerate}
The home CBG of an individual is determined by identifying the most frequently visited building within a census block group during nighttime hours. While these data cover a broad swath of the U.S. population, they may underrepresent populations with limited smartphone access or regions with inconsistent location signal coverage \cite{griffin2020mitigating}. Specifically, previous scholarship has examined foot-traffic data and unveiled sampling biases in specific demographics, such as Hispanic communities, economically disadvantaged households, and those with lower levels of education \cite{li2024understanding}. Notably, while the dataset encompasses individuals aged 65 and above, it intentionally omits those below 16 years of age` to protect the privacy of adolescents and children. 


For the dataset published and described in this paper, we calculated and provide weekly values for each NAICS (North American Industrial Classification System) code of the Point of Interest (POI) count, along with statistical summaries for three key metrics: total number of visitors, dwell time of visitors, and travel distance from their home census block group. For each NAICS code, we report the mean, standard deviation, minimum, 25th percentile, median, 75th percentile, and maximum values for these metrics. These calculations allow us to analyze and compare mobility patterns across different industries and sectors, providing insights into how various sectors were affected by the COVID-19 pandemic and subsequent lockdown measures.

\subsection*{Dataset Generation Scripts}

To obtain the dataset, we utilized weekly foot traffic data from Advan Data~\cite{advanresearch}. Key features extracted included the number of visitors, distance traveled from home, and dwell time for each point of interest (POI), along with the NAICS code corresponding to each POI. After preprocessing the raw downloaded data, we consolidated all weekly records into a single data frame containing the following fields: placekey, naics\_code, raw\_visit\_counts, raw\_visitor\_counts, distance\_from\_home, and median\_dwell. Subsequently, we grouped the NAICS codes at the 2-digit, 4-digit, and 6-digit levels, aggregating them based on key metrics such as visitor counts. We then computed descriptive statistics for each metric before performing post-processing steps to structure and store the data according to the format outlined in the Data Records section. The scripts for generating and formatting the dataset can be found in the GitHub repository under \textit{data.py} and \textit{dataset.ipynb}, respectively. Notably, these scripts assume that the data have been downloaded and preprocessed locally, as API implementation and data retrieval from Dewey Data are publicly available on their website, but accessing the data requires a paid subscription.







\section*{Data Records}
In this data descriptor paper, we provide comprehensive datasets for each specified metric (distance, dwell time, and visitors), structured according to NAICS codes. The dataset comprises 20 files corresponding to two-digit NAICS codes and an additional 174 files corresponding to four-digit NAICS codes. Each file includes weekly statistical summaries for its respective NAICS category, covering the 208 weeks period from January 7, 2019, to January 2, 2023. These summaries include count, mean, standard deviation (std), minimum (min), first quartile (q1), median (q2), third quartile (q3), maximum (max), and total sum.

Table \ref{tab:example_combined_11} displays an excerpt of our dataset (first and last four weeks) for NAICS Code 11 (Agriculture, Forestry, Fishing, and Hunting). In the first week observed (01/07/2019), 607 unique POIs were visited with distance information linked. Visitors to these POIs traveled an average of 28,064 meters from their home CBG and stayed approximately 168 minutes. Over time, the number of observed POIs in NAICS 11 decreased, likely reflecting both changes in user coverage and potential shifts in mobility patterns. Frequently, more than half of the POIs in this sector received four or fewer weekly visits, indicating relatively low, localized foot traffic.

The dataset~\cite{osf} can be downloaded from OSF at \url{https://doi.org/10.17605/OSF.IO/BDFGZ}.

\begin{table*}
    \centering
   
    \caption{Data examples for NAICS Code 11 (Agriculture, Forestry, Fishing, and Hunting) including distance traveled, dwell time (in minutes), and number of visitors for the available data (2019-2023)}
    \label{tab:example_combined_11}
    \begin{tabular}{|l|r|r|r|r|l|r|r|}
    \hline
    metrics & 2019-01-07 & 2019-01-14 & 2019-01-21 & 2019-01-28 & ... & 2022-12-19 & 2022-12-26  \\
    \multicolumn{8}{|l|}{\textbf{Distance Traveled [data/distance/11.csv]}} \\
    count & 607 & 601 & 611 & 595 & ... & 546 & 513  \\
    mean & 61088.98 & 36473.07 & 37047.23 & 49446.97 & ... & 77743.15 & 57229.75  \\
    std & 287133.84 & 100922.37 & 126140.35 & 157414.60 & ... & 350132.93 & 188517.03  \\
    min & 3 & 3 & 3 & 3 & ... & 26 & 21  \\
    q1 & 6452 & 5773 & 6760.50 & 6445 & ... & 5967.50 & 5925  \\
    q2 & 11431 & 11211 & 12259 & 11712 & ... & 12318 & 12855  \\
    q3 & 22674 & 23609 & 24434.50 & 24681 & ... & 25307 & 28293  \\
    max & 3831387 & 1070736 & 2367695 & 1606120 & ... & 3838216 & 1967350  \\
    sum & 37081012 &21920318 &22635857 &29420949 &... & 42447759 & 29358860 \\

    \multicolumn{8}{|l|}{\textbf{Dwell Time (Minutes) [data/dwell/11.csv]}} \\
    count & 597 & 604 & 617 & 602 & ... & 579 & 539  \\
    mean & 167.90 & 173.30 & 180.07 & 169.71 & ... & 155.31 & 142.06  \\
    std & 242.70 & 259.59 & 269.71 & 262.16 & ... & 251.33 & 241.58  \\
    min & 1 & 1 & 1 & 1 & ... & 1 & 1  \\
    q1 & 9 & 9 & 9 & 8 & ... & 4 & 4  \\
    q2 & 64 & 64 & 56 & 58.50 & ... & 36 & 33  \\
    q3 & 231 & 243.25 & 263 & 234.50 & ... & 233 & 185 \\
    max & 1399 & 1936 & 1440 & 1440 & ... & 1418 & 1414  \\
    sum & 100236 & 104674 & 111106 & 102165 & ... & 89926 & 76569 \\

    \multicolumn{8}{|l|}{\textbf{Number of Visitors [data/visitors/11.csv]}} \\
    count & 791 & 791 & 809 & 776 & ... & 761 & 703  \\
    mean & 252.01 & 253.08 & 252.72 & 249.30 & ... & 74.37 & 80.28  \\
    std & 4191.65 & 4262.23 & 4261.72 & 4109.67 & ... & 367.70 & 410.09  \\
    min & 1 & 1 & 1 & 1 & ... & 1 & 1  \\
    q1 & 2 & 2 & 2 & 2 & ... & 1 & 1  \\
    q2 & 6 & 6 & 5 & 6 & ... & 4 & 4  \\
    q3 & 25.50 & 26 & 27 & 28 & ... & 20 & 21.50  \\
    max & 117400 & 119401 & 120723 & 114031 & ... & 5438 & 7336  \\
    sum & 199342 & 200183 & 204451 & 193453 & ... & 56595 & 56439 \\
    \hline
    \end{tabular}
    \end{table*}

\section*{Technical Validation}

 In interpreting these data, it is important to recognize certain limitations. Prior research has identified sampling biases in mobile device data, particularly underrepresentation of certain demographic groups such as Hispanic populations and individuals from lower socioeconomic backgrounds. While these biases are inherent in mobility datasets, we believe the large sample size and diversity of points of interest mitigate, but do not eliminate, their effects.

\subsection*{Consistency with Other Sources}

\begin{figure}[t]
    \centering
    \includegraphics[width=0.99\linewidth]{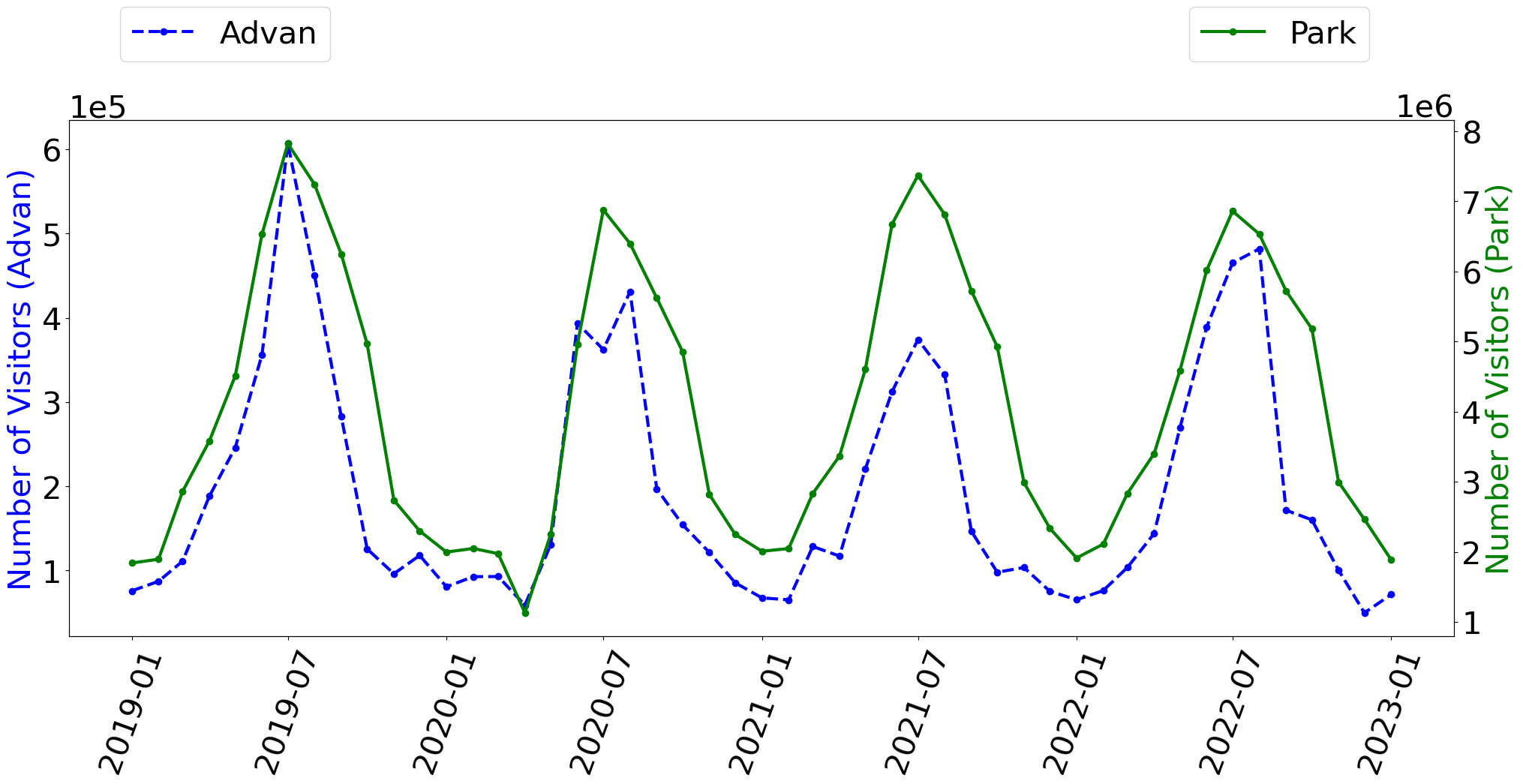}
    \caption{Comparison of visitor statistics for the Top-10 largest National Parks in the United States using different data sources, ADVAN Data \cite{advanresearch} and Park Visitor Data \cite{nps} }
     \label{fig:park_data}
\end{figure}

Data obtained from National Parks in the United States~\cite{nps} is used to assess the consistency of the provided dataset. We captured data for the top 10 most visited parks, as shown in Table \ref{tab:national_parks}, and extracted the corresponding data from the Advan dataset. The results are presented in Figure \ref{fig:park_data}. The left figure displays data from the National Parks, while the right figure presents data from the Advan dataset, which we used to generate the dataset. The charts illustrate the number of visitors from January 2019 to January 2023 in both the National Parks and Advan datasets. As shown, the patterns in both datasets are similar, with higher visitation during the summer and lower visitation in the winter. Both charts also capture the impact of the COVID-19 outbreak. The two time series exhibit a high positive correlation having a Pearson correlation of $0.875$. However, the absolute visitor counts in the National Parks dataset are greater than those in the Advan data. This is because every visit is recorded National Park Dataset. In contrast, the Advan dataset does not capture all visits, as it only includes a sample of visitors using applications tracked by Advan. We note that this comparison is qualitative in nature, based on visual inspection of trends rather than direct one-to-one statistical validation.

\begin{table}
    \centering
    \caption{Top Visited Locations from National Parks Data with Their Representative Coordinates in Advan Dataset}
    \begin{tabular}{|l|r|r|}
        \hline
        Name & Latitude & Longitude \\
        \hline
        Stonewall NM & 40.7336916 & -74.0019957 \\
        Golden Gate NRA & 37.848086 & -122.529527 \\
        Great Smoky Mountains NP & 35.62186 & -83.49738 \\
        Yellowstone NP & 44.46450 & -110.63897 \\
        Acadia NP & 44.37460 & -68.25935 \\
        Glen Canyon NRA & 37.65611 & -111.09285 \\
        Yosemite NP & 37.73108 & -119.63837 \\
        Rocky Mountain NP & 40.31196 & -105.64602 \\
        Cape Cod NS & 41.94381 & -70.07656 \\
        Lincoln Memorial & 38.889222 & -77.050635 \\
        \hline
    \end{tabular}
    \label{tab:national_parks}
\end{table}

\begin{table*}
    \caption{Description of NAICS Code Sample used in our Technical Validation}
    \centering
    \begin{tabular}{|c|c|c|c|}
    \hline
         NAICS Code &   Description                     &   NAICS Code  &   Description                         \\\hline
        31-33       &   Manufacturing                   &   61          &   Education                           \\
        44-45       &   Retail                          &   62          &   Healthcare and Social Assistance    \\
        48-49       &   Transportation and Warehousing  &   71          &   Arts, Entertainment, Recreation     \\
        52          &   Financial and Insurance         &   72          &   Accommodation, Food Services        \\
        53          &   Real Estate, Rental, Leasing    &   82          &   Other Services                      \\
    \hline
    \end{tabular}
    \label{tab:naics_code_map}
\end{table*}

\begin{figure}[t]
    \centering
    \includegraphics[width=0.99\linewidth]{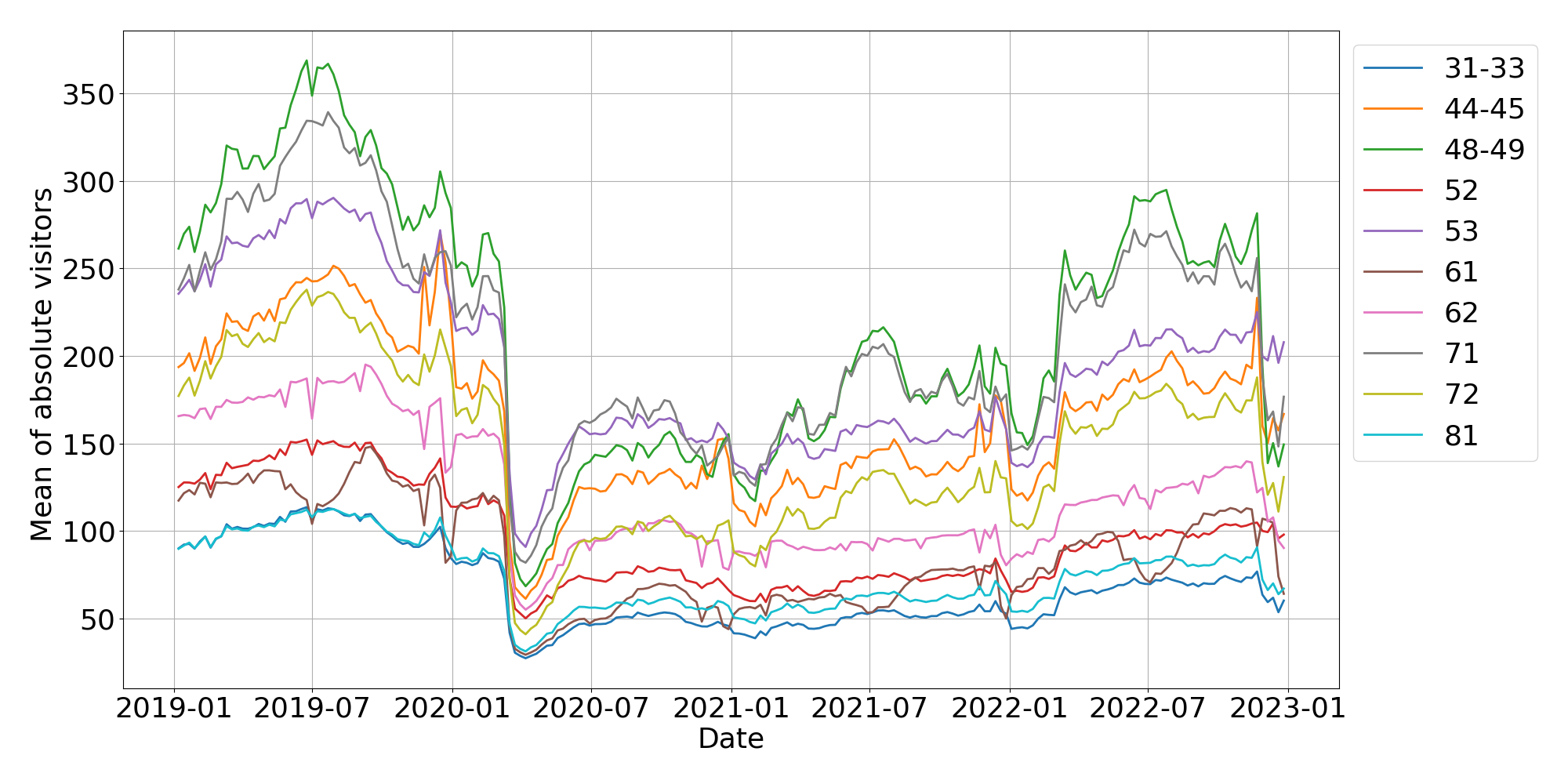}
    \caption{Visualization of the mean number of visitors for different NAICS sectors for the available data(2019-2023). Visitor counts show clear seasonality and large declines across most sectors during the COVID-19 pandemic in early 2020.}
    \label{fig:visitors-mean}
\end{figure}

\begin{figure}[t]
    \centering
    \includegraphics[width=0.99\linewidth]{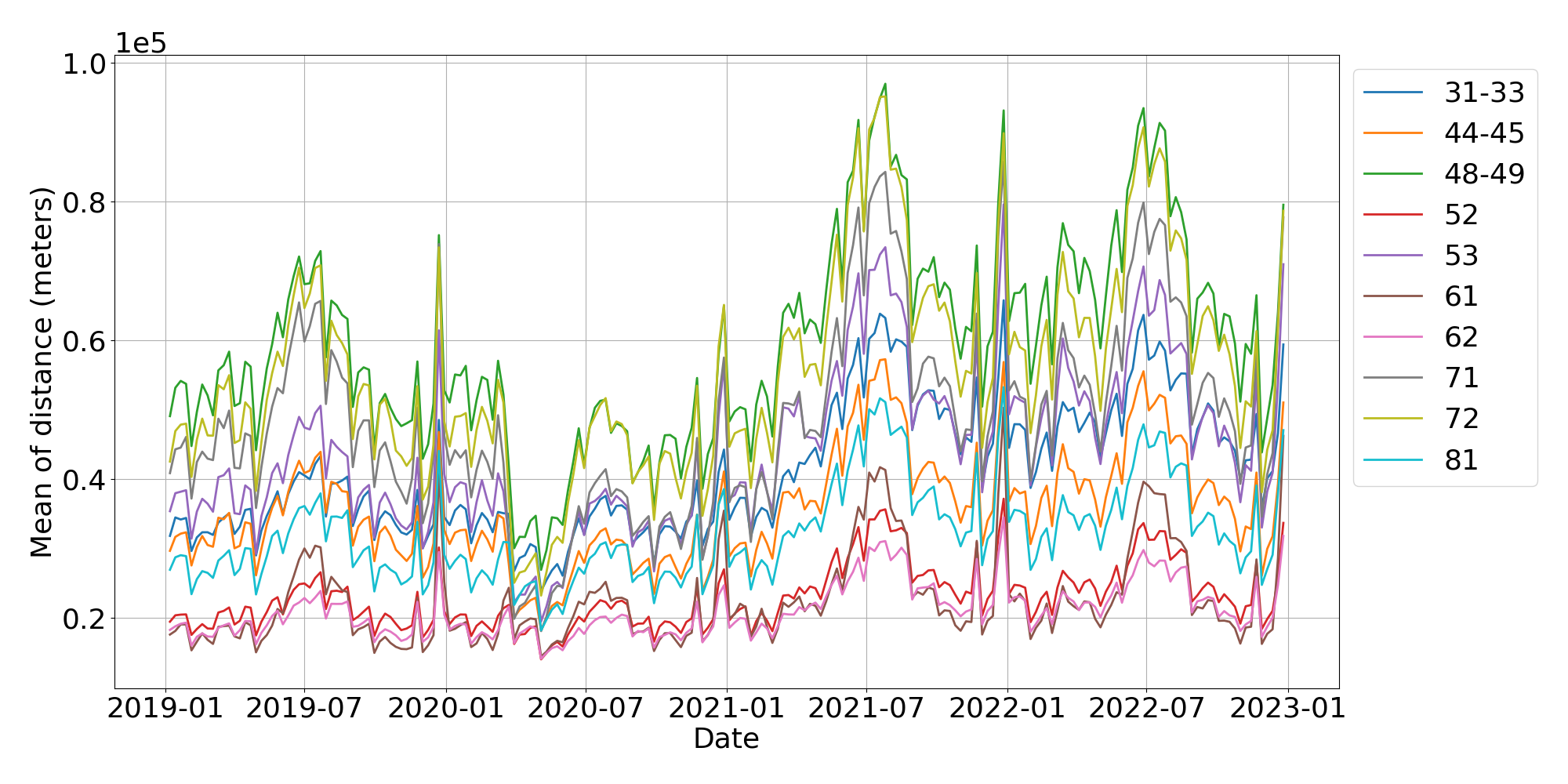}
    \caption{Visualization of the mean distance traveled for different NAICS sectors for the available data(2019-2023)}
    \label{fig:distance-mean}
\end{figure}

\begin{figure}[t]
    \centering
    \includegraphics[width=0.99\linewidth]{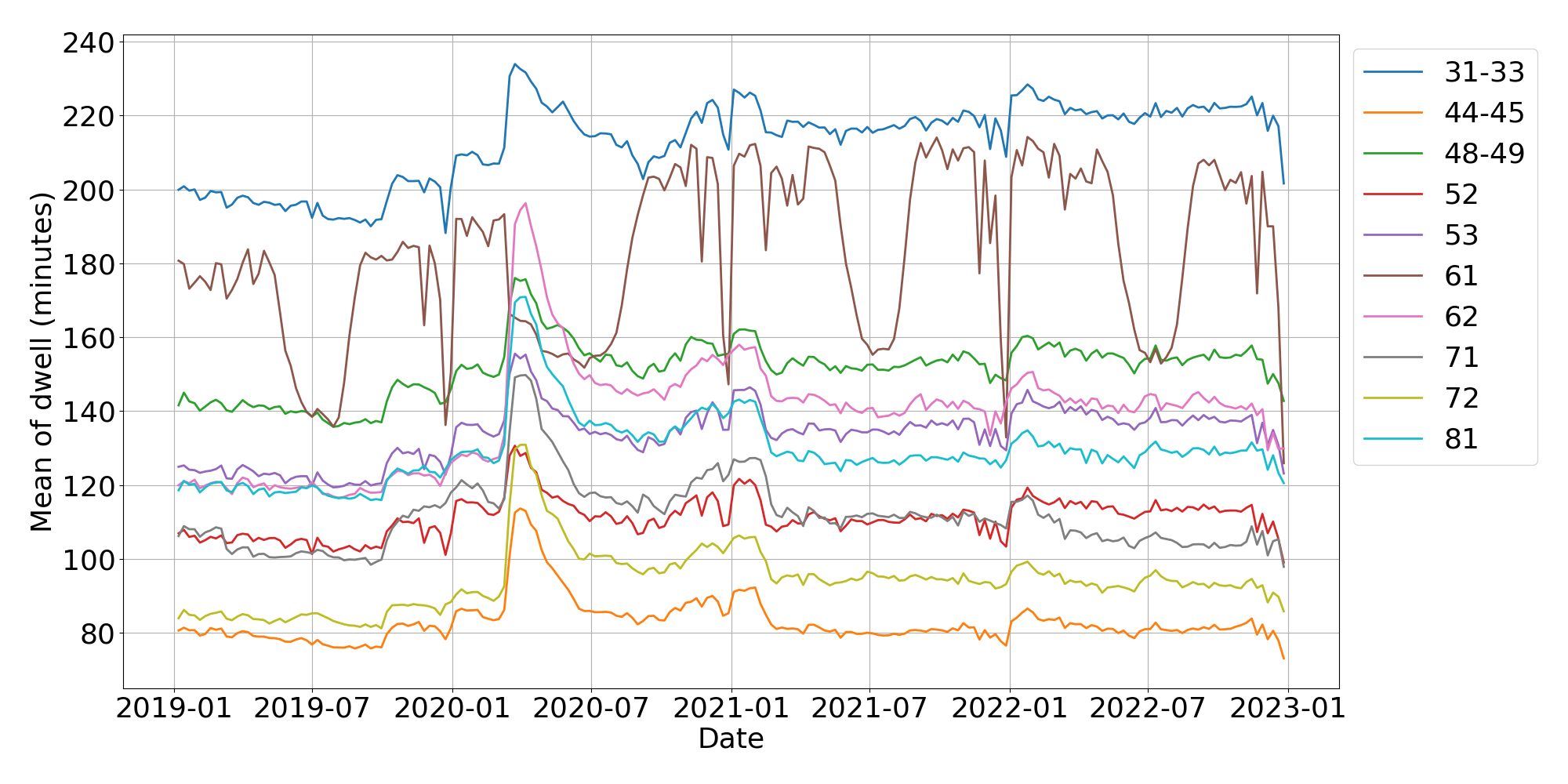}
    \caption{Visualization of the mean dwell time for different NAICS sectors for the available data(2019-2023)}
    \label{fig:dwell-mean}
\end{figure}
\subsection*{Temporal Analysis}
To demonstrate the breadth of the dataset, we focus on the ten NAICS sectors with the highest average POI counts. These sectors represent a diverse range of industries and provide a comprehensive overview of mobility patterns across different economic sectors. Figure~\ref{fig:visitors-mean} presents the average weekly number of visitors across these sectors from January 2019 to January 2023, revealing both seasonal fluctuations and the influence of the COVID-19 pandemic. The corresponding sectors for the ten two-digit NAICS Code  used in this analysis can be found in Table~\ref{tab:naics_code_map}.
For instance, NAICS 61 (Educational Services) shows marked dips during the traditional summer break and winter holidays, illustrating the expected academic calendar effect. By contrast, during the COVID-19 pandemic in 2020, visits in Sector 61 dropped to near zero, reflecting widespread school closures. By contrast, NAICS Sector 44-45 (Retail Trade) shows smaller declines, indicating that retail activities remained partially operational throughout much of the pandemic. Studying the detailed 4-digit NAICS sectors that we provide may help identify which human activities continued (legally or not) during the COVID-19 pandemic. This understanding may help decision-making to prevent future pandemics.

Figure~\ref{fig:distance-mean} displays the average travel distance by week for the same ten sectors, offering insights into the spatial extent of visitor movement. We observe that NAICS Sectors 71, 72, and 48-49 show an average travel distance approaching ~100 km, significantly higher than other sectors, a pattern likely driven by visits to hotels and related travel-oriented venues. We also observe that NAICS Sectors 52, 61, and 62 exhibit the smallest travel distances with an average of around 20 km. With the onset of the COVID-19 pandemic, average travel distances across sectors plummeted during April 2020 due to widespread business closures and travel restrictions but rebounded by June 2020, potentially reflecting loosening regulations and renewed demand for travel similar to what we observe for the number of visitors shown in Figure~\ref{fig:visitors-mean}

Finally, Figure~\ref{fig:dwell-mean} examines dwell time trends. This graph elucidates how visitors' engagement with different establishments evolves over time, shedding light on factors influencing the duration of stay, such as the nature of the venue or external events. While most sectors exhibit relatively stable dwell times trends year-round, NAICS 44-45 (Retail Trade) consistently shows lower dwell durations than other sectors, likely reflecting the nature of retail visits. In contrast, NAICS Sector 61 (Educational Services) experiences notable drops during summer periods as well as before the winter holidays period as schools close earlier. Taken together, these observations underscore the value of high-frequency mobility data for understanding how different sectors adapt to changing conditions and external disruptions.

\begin{table*}
    \centering
    \caption{Weekly average visitor counts by NAICS sector for sectors 11 - 33, plus correlations of mean visitors, distance traveled, and dwell time for each 4-digit subsector relative to its 2-digit parent sector.
Correlation Coefficients by the strength of correlation with the 2-digit sector above 0.9 are in bold, and from 0.7 to 0.9 are in italics.}
\label{tab:correlation_table1}
    \begin{tabular}{ccccc}
        \toprule
        NAICS & Mean Visitor Count & Mean Travel Distance & Mean Visitors & Mean Dwell Time \\
        \midrule
        11    & 209.6 &    &    &    \\
        1114  & 307.0 & {\CorrCell{0.925}} & {\CorrCell{0.994}} & {\CorrCell{0.943}} \\
        1151  & 11.9  & {\CorrCell{0.563}} & {\CorrCell{0.332}} & {\CorrCell{0.349}} \\
         & & & & \\
        21    & 118.7 &    &    &    \\
        2111  & 44.5  & {\CorrCell{0.576}} & {\CorrCell{0.179}} & {\CorrCell{0.331}} \\
        2131  & 176.7 & {\CorrCell{0.835}} & {\CorrCell{0.986}} & {\CorrCell{0.834}} \\
              &   &   &   &   \\
        22    & 102.9 &    &    &    \\
        2211  & 108.5 & {\CorrCell{0.998}} & {\CorrCell{0.999}} & {\CorrCell{0.997}} \\
        2212  & 30.8  & {\CorrCell{0.515}} & {\CorrCell{0.911}} & {\CorrCell{0.277}} \\
              &   &   &   &   \\
        23    & 56.0  &    &    &    \\
        2361  & 36.1  & {\CorrCell{0.763}} & {\CorrCell{0.968}} & {\CorrCell{0.911}} \\
        2371  & 31.6  & {\CorrCell{0.841}} & {\CorrCell{0.974}} & {\CorrCell{0.912}} \\
        2373  & 374.2 & {\CorrCell{0.200}} & {\CorrCell{0.257}} & {\CorrCell{-0.404}} \\
        2381  & 34.3  & {\CorrCell{0.953}} & {\CorrCell{0.968}} & {\CorrCell{0.982}} \\
        2382  & 52.9  & {\CorrCell{0.995}} & {\CorrCell{0.999}} & {\CorrCell{0.998}} \\
        2389  & 71.9  & {\CorrCell{0.995}} & {\CorrCell{0.996}} & {\CorrCell{0.991}} \\
              &   &   &   &   \\
        31-33 & 66.9  &    &    &    \\
        3118  & 128.6 & {\CorrCell{0.949}} & {\CorrCell{0.971}} & {\CorrCell{0.569}} \\
        3119  & 97.9  & {\CorrCell{0.415}} & {\CorrCell{0.972}} & {\CorrCell{0.651}} \\
        3121  & 65.4  & {\CorrCell{0.913}} & {\CorrCell{0.964}} & {\CorrCell{0.504}} \\
        3159  & 84.1  & {\CorrCell{0.655}} & {\CorrCell{0.971}} & {\CorrCell{0.730}} \\
        3162  & 74.2  & {\CorrCell{0.563}} & {\CorrCell{0.944}} & {\CorrCell{0.298}} \\
        3169  & 636.7 & {\CorrCell{0.501}} & {\CorrCell{0.912}} & {\CorrCell{0.212}} \\
        3219  & 38.8  & {\CorrCell{0.778}} & {\CorrCell{0.955}} & {\CorrCell{0.243}} \\
        3222  & 323.0 & {\CorrCell{0.520}} & {\CorrCell{0.992}} & {\CorrCell{0.800}} \\
        3231  & 80.7  & {\CorrCell{0.985}} & {\CorrCell{0.998}} & {\CorrCell{0.937}} \\
        3241  & 15.4  & {\CorrCell{0.186}} & {\CorrCell{0.613}} & {\CorrCell{0.302}} \\
        3255  & 27.5  & {\CorrCell{-0.039}}& {\CorrCell{0.157}} & {\CorrCell{0.485}} \\
        3261  & 18.5  & {\CorrCell{-0.020}}& {\CorrCell{0.680}} & {\CorrCell{0.217}} \\
        3271  & 22.5  & {\CorrCell{0.580}} & {\CorrCell{0.835}} & {\CorrCell{0.722}} \\
        3272  & 70.8  & {\CorrCell{0.918}} & {\CorrCell{0.993}} & {\CorrCell{0.716}} \\
        3312  & 18.3  & {\CorrCell{0.544}} & {\CorrCell{0.778}} & {\CorrCell{0.529}} \\
        3313  & 85.4  & {\CorrCell{0.129}} & {\CorrCell{0.937}} & {\CorrCell{0.211}} \\
        3323  & 37.0  & {\CorrCell{0.372}} & {\CorrCell{0.779}} & {\CorrCell{0.442}} \\
        3325  & 25.0  & {\CorrCell{0.301}} & {\CorrCell{0.788}} & {\CorrCell{0.553}} \\
        3328  & 71.0  & {\CorrCell{0.863}} & {\CorrCell{0.961}} & {\CorrCell{0.908}} \\
        3329  & 272.9 & {\CorrCell{0.202}} & {\CorrCell{0.809}} & {\CorrCell{0.131}} \\
        3331  & 22.8  & {\CorrCell{0.227}} & {\CorrCell{0.667}} & {\CorrCell{0.573}} \\
        3334  & 7.0   & {\CorrCell{0.003}} & {\CorrCell{0.820}} & {\CorrCell{-0.232}} \\
        3344  & 229.4 & {\CorrCell{0.213}} & {\CorrCell{0.918}} & {\CorrCell{0.198}} \\
        3345  & 978.9 & {\CorrCell{0.548}} & {\CorrCell{0.902}} & {\CorrCell{0.285}} \\
        3352  & 46.3  & {\CorrCell{0.889}} & {\CorrCell{0.991}} & {\CorrCell{0.856}} \\
        3361  & 93.0  & {\CorrCell{-0.024}}& {\CorrCell{0.782}} & {\CorrCell{0.287}} \\
        3364  & 550.0 & {\CorrCell{0.293}} & {\CorrCell{0.925}} & {\CorrCell{0.465}} \\
        3369  & 1820.9 & {\CorrCell{0.325}} & {\CorrCell{0.850}} & {\CorrCell{0.074}} \\
        3391  & 17.0  & {\CorrCell{0.206}} & {\CorrCell{0.853}} & {\CorrCell{0.584}} \\
        3399  & 38.5  & {\CorrCell{0.977}} & {\CorrCell{0.990}} & {\CorrCell{0.896}} \\
        \bottomrule
        \end{tabular}
        
\end{table*}

\begin{table*}
    \centering

    \caption{Weekly average visitor counts by NAICS sector for sectors 42 - 49, plus correlations of mean visitors, distance traveled, and dwell time for each 4-digit subsector relative to its 2-digit parent sector.
Correlation Coefficients by the strength of correlation with the 2-digit sector above 0.9 are in bold, and from 0.7 to 0.9 are in italics.}
\label{tab:correlation_table4}
    \begin{tabular}{ccccc}
        \toprule
        NAICS & Mean Visitor Count & Mean Travel Distance & Mean Visitors & Mean Dwell Time \\
        \midrule
        42-43 & 65.2 &  &  &  \\
        4231  & 20.4 & \CorrCell{0.591} & \CorrCell{0.819} & \CorrCell{0.767} \\
        4232  & 10.8 & \CorrCell{-0.038} & \CorrCell{0.399} & \CorrCell{-0.117} \\
        4233  & 20.0 & \CorrCell{0.769} & \CorrCell{0.825} & \CorrCell{0.846} \\
        4234  & 61.6 & \CorrCell{0.700} & \CorrCell{0.897} & \CorrCell{0.841} \\
        4235  & 22.7 & \CorrCell{0.551} & \CorrCell{0.765} & \CorrCell{0.159} \\
        4236  & 24.6 & \CorrCell{0.834} & \CorrCell{0.902} & \CorrCell{0.710} \\
        4237  & 51.5 & \CorrCell{0.977} & \CorrCell{0.996} & \CorrCell{0.861} \\
        4238  & 81.1 & \CorrCell{0.988} & \CorrCell{0.999} & \CorrCell{0.977} \\
        4242  & 45.8 & \CorrCell{0.935} & \CorrCell{0.976} & \CorrCell{0.432} \\
        4244  & 140.2 & \CorrCell{0.581} & \CorrCell{0.969} & \CorrCell{0.448} \\
        4245  & 7.7 & \CorrCell{0.217} & \CorrCell{0.503} & \CorrCell{0.116} \\
        4246  & 10.7 & \CorrCell{0.775} & \CorrCell{0.747} & \CorrCell{0.405} \\
        4247  & 104.3 & \CorrCell{0.490} & \CorrCell{0.544} & \CorrCell{-0.013} \\
        4249  & 54.4 & \CorrCell{0.223} & \CorrCell{0.016} & \CorrCell{0.317} \\
              &   &   &   &   \\
        44-45 & 165.3 &  &  &  \\
        4411  & 70.6  & \CorrCell{0.949} & \CorrCell{0.844} & \CorrCell{0.893} \\
        4412  & 43.6  & \CorrCell{0.964} & \CorrCell{0.925} & \CorrCell{0.543} \\
        4421  & 95.3  & \CorrCell{0.961} & \CorrCell{0.987} & \CorrCell{0.938} \\
        4431  & 104.0 & \CorrCell{0.958} & \CorrCell{0.984} & \CorrCell{0.868} \\
        4441  & 69.1  & \CorrCell{0.966} & \CorrCell{0.737} & \CorrCell{0.876} \\
        4442  & 39.2  & \CorrCell{0.933} & \CorrCell{0.958} & \CorrCell{0.384} \\
        4451  & 121.6 & \CorrCell{0.974} & \CorrCell{0.942} & \CorrCell{0.918} \\
        4453  & 81.7  & \CorrCell{0.986} & \CorrCell{0.978} & \CorrCell{0.919} \\
        4461  & 249.8 & \CorrCell{0.978} & \CorrCell{0.986} & \CorrCell{0.967} \\
        4471  & 183.0 & \CorrCell{0.830} & \CorrCell{0.867} & \CorrCell{0.560} \\
        4482  & 389.3 & \CorrCell{0.975} & \CorrCell{0.980} & \CorrCell{0.945} \\
        4483  & 406.8 & \CorrCell{0.966} & \CorrCell{0.991} & \CorrCell{0.972} \\
        4511  & 133.8 & \CorrCell{0.989} & \CorrCell{0.995} & \CorrCell{0.987} \\
        4522  & 293.6 & \CorrCell{0.976} & \CorrCell{0.954} & \CorrCell{0.889} \\
        4523  & 248.4 & \CorrCell{0.790} & \CorrCell{0.904} & \CorrCell{0.620} \\
        4531  & 72.8  & \CorrCell{0.988} & \CorrCell{0.978} & \CorrCell{0.942} \\
        4532  & 166.7 & \CorrCell{0.988} & \CorrCell{0.993} & \CorrCell{0.972} \\
        4533  & 60.1  & \CorrCell{0.975} & \CorrCell{0.967} & \CorrCell{0.959} \\
        4543  & 27.5  & \CorrCell{0.743} & \CorrCell{0.864} & \CorrCell{0.260} \\
              &   &   &   &   \\
        48-49 & 218.6 &  &  &  \\
        4811  & 2923.4 & \CorrCell{0.877} & \CorrCell{0.990} & \CorrCell{0.623} \\
        4821  & 447.4   & \CorrCell{0.954} & \CorrCell{0.993} & \CorrCell{0.849} \\
        4831  & 518.8   & \CorrCell{0.391} & \CorrCell{0.886} & \CorrCell{0.094} \\
        4841  & 14.7    & \CorrCell{0.183} & \CorrCell{0.577} & \CorrCell{0.324} \\
        4851  & 42.9    & \CorrCell{0.349} & \CorrCell{0.825} & \CorrCell{0.585} \\
        4852  & 234.9   & \CorrCell{0.838} & \CorrCell{0.992} & \CorrCell{0.821} \\
        4853  & 169.7   & \CorrCell{0.961} & \CorrCell{0.999} & \CorrCell{0.964} \\
        4854  & 7.6     & \CorrCell{0.044} & \CorrCell{0.656} & \CorrCell{0.176} \\
        4859  & 437.4   & \CorrCell{0.933} & \CorrCell{0.993} & \CorrCell{0.892} \\
        4871  & 135.4   & \CorrCell{0.882} & \CorrCell{0.820} & \CorrCell{0.270} \\
        4881  & 1194.9  & \CorrCell{0.952} & \CorrCell{0.991} & \CorrCell{0.774} \\
        4883  & 1903.2  & \CorrCell{0.720} & \CorrCell{0.908} & \CorrCell{0.313} \\
        4884  & 99.0    & \CorrCell{0.946} & \CorrCell{0.971} & \CorrCell{0.963} \\
        4885  & 18.2    & \CorrCell{0.502} & \CorrCell{0.700} & \CorrCell{0.517} \\
        4889  & 41.9    & \CorrCell{0.400} & \CorrCell{0.855} & \CorrCell{0.365} \\
        4911  & 24.7    & \CorrCell{0.732} & \CorrCell{0.833} & \CorrCell{0.230} \\
        4921  & 111.6   & \CorrCell{0.955} & \CorrCell{0.958} & \CorrCell{0.956} \\
        4931  & 95.9    & \CorrCell{0.778} & \CorrCell{0.815} & \CorrCell{0.624} \\
        \bottomrule
        \end{tabular}
        
\end{table*}

\begin{table*}
    \centering
    \caption{Weekly average visitor counts by NAICS sector for sectors 51 - 61, plus correlations of mean visitors, distance traveled, and dwell time for each 4-digit subsector relative to its 2-digit parent sector.
Correlation Coefficients by the strength of correlation with the 2-digit sector above 0.9 are in bold, and from 0.7 to 0.9 are in italics.}
\label{tab:correlation_table5}
    \begin{tabular}{ccccc}
        \toprule
        NAICS & Mean Visitor Count & Mean Travel Distance & Mean Visitors & Mean Dwell Time \\
        \midrule
        51    & 136.3 &      &    &    \\
        5111  & 21.1  & \CorrCell{0.285} & \CorrCell{0.745} & \CorrCell{0.018} \\
        5121  & 185.7 & \CorrCell{0.964} & \CorrCell{0.975} & \CorrCell{0.887} \\
        5122  & 108.8 & \CorrCell{0.908} & \CorrCell{0.982} & \CorrCell{0.914} \\
        5151  & 113.5 & \CorrCell{0.976} & \CorrCell{0.980} & \CorrCell{0.765} \\
        5152  & 149.4 & \CorrCell{0.737} & \CorrCell{0.987} & \CorrCell{0.877} \\
        5179  & 138.7 & \CorrCell{0.240} & \CorrCell{0.919} & \CorrCell{0.329} \\
        5182  & 94.0  & \CorrCell{0.775} & \CorrCell{0.926} & \CorrCell{0.529} \\
        5191  & 46.5  & \CorrCell{0.932} & \CorrCell{0.939} & \CorrCell{0.909} \\
              &   &   &   &   \\
        52    & 95.0       &     &    \\
        5221  & 100.0 & \CorrCell{0.981} & \CorrCell{0.993} & \CorrCell{0.673} \\
        5222  & 56.0  & \CorrCell{0.960} & \CorrCell{0.996} & \CorrCell{0.851} \\
        5223  & 137.6 & \CorrCell{0.993} & \CorrCell{0.990} & \CorrCell{0.984} \\
        5231  & 55.7  & \CorrCell{0.805} & \CorrCell{0.987} & \CorrCell{0.902} \\
        5241  & 113.5 & \CorrCell{0.911} & \CorrCell{0.998} & \CorrCell{0.844} \\
        5242  & 54.3  & \CorrCell{0.994} & \CorrCell{0.991} & \CorrCell{0.925} \\
        5259  & 176.5 & \CorrCell{0.597} & \CorrCell{0.977} & \CorrCell{0.178} \\
              &   &   &   &   \\
        53    & 192.4 &      &    &    \\
        5311  & 592.0 & \CorrCell{0.940} & \CorrCell{0.982} & \CorrCell{0.078} \\
        5312  & 68.5  & \CorrCell{0.993} & \CorrCell{0.977} & \CorrCell{0.990} \\
        5321  & 306.8 & \CorrCell{0.971} & \CorrCell{0.903} & \CorrCell{0.849} \\
        5322  & 95.3  & \CorrCell{0.982} & \CorrCell{0.990} & \CorrCell{0.917} \\
        5323  & 22.6  & \CorrCell{0.630} & \CorrCell{0.936} & \CorrCell{0.829} \\
        5324  & 133.0 & \CorrCell{0.864} & \CorrCell{0.616} & \CorrCell{0.721} \\
              &   &   &   &   \\
        54    & 103.5 &     &    &     \\
        5411  & 73.3  & \CorrCell{0.919} & \CorrCell{0.994} & \CorrCell{0.955} \\
        5412  & 106.3 & \CorrCell{0.994} & \CorrCell{0.999} & \CorrCell{0.972} \\
        5413  & 64.2  & \CorrCell{0.817} & \CorrCell{0.996} & \CorrCell{0.898} \\
        5416  & 148.4 & \CorrCell{0.984} & \CorrCell{0.998} & \CorrCell{0.748} \\
        5417  & 79.1  & \CorrCell{0.413} & \CorrCell{0.958} & \CorrCell{0.007} \\
        5418  & 131.8 & \CorrCell{0.986} & \CorrCell{0.999} & \CorrCell{0.935} \\
              &   &   &   &   \\
        56    & 126.5 &      &      &      \\
        5613  & 107.6 & \CorrCell{0.854} & \CorrCell{0.994} & \CorrCell{0.913} \\
        5614  & 220.9 & \CorrCell{0.637} & \CorrCell{0.943} & \CorrCell{0.691} \\
        5616  & 149.2 & \CorrCell{0.957} & \CorrCell{0.997} & \CorrCell{0.966} \\
        5619  & 125.9 & \CorrCell{0.615} & \CorrCell{0.908} & \CorrCell{0.551} \\
        5621  & 35.3  & \CorrCell{0.586} & \CorrCell{0.970} & \CorrCell{0.725} \\
        5622  & 31.1  & \CorrCell{0.814} & \CorrCell{0.968} & \CorrCell{0.678} \\
        5629  & 25.3  & \CorrCell{0.773} & \CorrCell{0.929} & \CorrCell{0.800} \\
              &   &   &   &   \\
        61    & 86.2  &     &     &     \\
        6111  & 52.5  & \CorrCell{0.953} & \CorrCell{0.920} & \CorrCell{0.982} \\
        6112  & 284.7 & \CorrCell{0.881} & \CorrCell{0.983} & \CorrCell{0.759} \\
        6113  & 258.3 & \CorrCell{0.934} & \CorrCell{0.982} & \CorrCell{0.922} \\
        6114  & 75.5  & \CorrCell{0.460} & \CorrCell{0.937} & \CorrCell{0.074} \\
        6115  & 122.8 & \CorrCell{0.879} & \CorrCell{0.941} & \CorrCell{0.612} \\
        6116  & 86.1  & \CorrCell{0.919} & \CorrCell{0.948} & \CorrCell{0.132} \\
        6117  & 330.4 & \CorrCell{0.331} & \CorrCell{0.947} & \CorrCell{0.607} \\
        \bottomrule
        \end{tabular}
        
\end{table*}


\begin{table*}
    \centering
    \caption{Weekly average visitor counts by NAICS sector for sectors 62 - 92, plus correlations of mean visitors, distance traveled, and dwell time for each 4-digit subsector relative to its 2-digit parent sector.
Correlation Coefficients by the strength of correlation with the 2-digit sector above 0.9 are in bold, and from 0.7 to 0.9 are in italics.}
    \label{tab:correlation_table7}
    \begin{tabular}{ccccc}
        \toprule
        NAICS & Mean Visitor Count & Mean Travel Distance & Mean Visitors & Mean Dwell Time \\
        \midrule
        62     & 121.2 &      &  &     \\
        6211   & 179.2 & \CorrCell{0.996} & \CorrCell{0.998} & \CorrCell{0.989} \\
        6212   & 62.6  & \CorrCell{0.976} & \CorrCell{0.992} & \CorrCell{0.987} \\
        6214   & 94.3  & \CorrCell{0.975} & \CorrCell{0.999} & \CorrCell{0.997} \\
        6215   & 104.0 & \CorrCell{0.968} & \CorrCell{0.996} & \CorrCell{0.973} \\
        6216   & 57.4  & \CorrCell{0.976} & \CorrCell{0.978} & \CorrCell{0.903} \\
        6219   & 82.9  & \CorrCell{0.897} & \CorrCell{0.995} & \CorrCell{0.968} \\
        6221   & 345.8 & \CorrCell{0.947} & \CorrCell{0.981} & \CorrCell{0.994} \\
        6222   & 73.7  & \CorrCell{0.818} & \CorrCell{0.992} & \CorrCell{0.869} \\
        6223   & 140.7 & \CorrCell{0.978} & \CorrCell{0.999} & \CorrCell{0.997} \\
        6231   & 33.8  & \CorrCell{0.899} & \CorrCell{0.976} & \CorrCell{0.892} \\
        6233   & 31.1  & \CorrCell{0.908} & \CorrCell{0.990} & \CorrCell{0.937} \\
        6241   & 59.0  & \CorrCell{0.891} & \CorrCell{0.981} & \CorrCell{0.931} \\
        6242   & 32.6  & \CorrCell{0.888} & \CorrCell{0.984} & \CorrCell{0.906} \\
        6244   & 27.5  & \CorrCell{0.944} & \CorrCell{0.980} & \CorrCell{0.707} \\
              &   &   &   &   \\
        71     & 211.4 &       &     &      \\
        7111   & 200.5 & \CorrCell{0.941} & \CorrCell{0.959} & \CorrCell{0.892} \\
        7112   & 489.1 & \CorrCell{0.913} & \CorrCell{0.898} & \CorrCell{-0.147} \\
        7113   & 184.2 & \CorrCell{0.981} & \CorrCell{0.962} & \CorrCell{0.891} \\
        7121   & 219.0 & \CorrCell{0.988} & \CorrCell{0.984} & \CorrCell{0.870} \\
        7131   & 933.1 & \CorrCell{0.986} & \CorrCell{0.966} & \CorrCell{0.819} \\
        7132   & 225.1 & \CorrCell{0.961} & \CorrCell{0.978} & \CorrCell{0.907} \\
              &   &   &   &   \\
        72     & 145.3 &     &      &     \\
        7211   & 102.8 & \CorrCell{0.990} & \CorrCell{0.976} & \CorrCell{0.435} \\
        7212   & 131.9 & \CorrCell{0.942} & \CorrCell{0.848} & \CorrCell{0.578} \\
        7223   & 108.4 & \CorrCell{0.960} & \CorrCell{0.992} & \CorrCell{0.962} \\
        7224   & 177.0 & \CorrCell{0.970} & \CorrCell{0.996} & \CorrCell{0.982} \\
              &   &   &   &   \\
        81     & 73.9  &      &     &     \\
        8111   & 44.5  & \CorrCell{0.982} & \CorrCell{0.983} & \CorrCell{0.931} \\
        8114   & 113.0 & \CorrCell{0.980} & \CorrCell{0.995} & \CorrCell{0.976} \\
        8122   & 116.7 & \CorrCell{0.871} & \CorrCell{0.519} & \CorrCell{0.809} \\
        8123   & 62.4  & \CorrCell{0.978} & \CorrCell{0.983} & \CorrCell{0.969} \\
        8131   & 27.1  & \CorrCell{0.980} & \CorrCell{0.977} & \CorrCell{0.983} \\
        8132   & 45.1  & \CorrCell{0.848} & \CorrCell{0.935} & \CorrCell{0.911} \\
        8133   & 117.1 & \CorrCell{0.962} & \CorrCell{0.983} & \CorrCell{0.962} \\
        8134   & 106.5 & \CorrCell{0.917} & \CorrCell{0.987} & \CorrCell{0.877} \\
        8139   & 25.0  & \CorrCell{0.207} & \CorrCell{0.915} & \CorrCell{0.422} \\
              &   &   &   &   \\
        92     & 77.6  &  &  &       \\
        9211   & 357.5 & \CorrCell{0.532} & \CorrCell{0.954} & \CorrCell{0.721} \\
        9221   & 65.3  & \CorrCell{0.995} & \CorrCell{0.999} & \CorrCell{0.993} \\
        9231   & 60.7  & \CorrCell{0.573} & \CorrCell{0.985} & \CorrCell{0.855} \\
        9261   & 113.3 & \CorrCell{0.967} & \CorrCell{0.990} & \CorrCell{0.779} \\
        9281   & 273.6 & \CorrCell{0.810} & \CorrCell{0.992} & \CorrCell{0.843} \\
        \bottomrule
        \end{tabular}

\end{table*}
\vspace{-0.3cm}
\subsection*{Correlation Analysis}

To assess how closely four-digit NAICS subsectors mirror their overarching two-digit sector, we computed Pearson’s correlation coefficients between each two-digit sector and its corresponding four-digit subsectors for three key variables: visitor count, travel distance, and dwell time. By quantifying these relationships, we can identify patterns of internal consistency or divergence within each sector, thereby informing both high-level and more granular analyses. The resulting correlation matrices are presented in Tables~\ref{tab:correlation_table1}–\ref{tab:correlation_table7}. 
Overall, most four-digit subsectors show correlations exceeding 0.9 with their two-digit aggregates for mean visitor counts, suggesting that visits generally follow consistent trends within a given sector. However, travel distance and dwell time often exhibit more variation. In many instances, one or two four-digit subsectors deviate notably from the overall sector trend. Notably, three two-digit sectors: 31–33 (Manufacturing), 42 (Wholesale), and 48–49 (Transportation and Warehousing)—display lower correlations with their four-digit subsectors for both travel distance and dwell time. These findings indicate that while sector-level analyses can be illustrative in many cases, examining more granular industry classifications can reveal important behavioral distinctions.

Collectively, these analyses yield a multifaceted overview of mobility patterns in the dataset, highlighting key similarities and differences within and across sectors. By integrating descriptive statistics, correlation measures, and visualizations, we not only strengthen the technical rigor of our dataset but also provide a rich foundation for future research. We anticipate that scholars in fields such as social science, activity-based intelligence, and transportation planning will find the complete dataset—spanning 208 weeks and covering 194 NAICS codes—highly useful for exploring the dynamics of human mobility across diverse contexts.

The data associated with this study are provided alongside this manuscript and organized by NAICS code and week. It is formatted to be compatible with commonly used statistical software packages, facilitating its integration into a wide range of analyses. By making these data openly available, we hope to encourage collaboration and spur further investigations into the mechanisms driving human mobility and their implications for public health, urban planning, and economic policy.

\vspace{-0.3cm}
\section*{Usage Notes}
\begin{enumerate}
    \item The provided data represent only an approximately~10\% sample of the total population. While this sample offers valuable insights into mobility patterns, researchers should be aware of potential biases inherent in any sampling process. These biases could arise from factors such as geographic location, socioeconomic status, or demographic characteristics, which may not be fully representative of the broader population.
    \item The data rely on individuals having access to and using cell phones, introducing a potential source of bias. Certain demographics or regions may have differential access to mobile technology. Researchers should be mindful of these limitations and consider how they may impact the generalizability of findings, particularly in studies focusing on vulnerable or underserved populations.
    \item The dataset primarily captures interactions with economic Points of Interest (POIs) categorized by the North American Industry Classification System (NAICS). While this provides valuable insights into economic activity and consumer behavior, it is not a comprehensive picture of individual mobility. Other aspects of mobility, such as recreational activities, social interactions, or healthcare access, may not be fully captured. Researchers should consider supplementing this data with additional sources to provide a more holistic understanding of human mobility patterns.
    \item When interpreting metrics such as dwell time, researchers should be mindful of the inherent variability in human behavior. Dwell time can encompass a wide range of activities, from prolonged stays (e.g., working a shift) to brief visits (e.g., making a quick purchase or pickup). Understanding the nuances and potential distribution of dwell time is essential for interpreting findings and avoiding misinterpretations or oversimplifications.
    
\end{enumerate}
\vspace{-0.3cm}
\section*{Code availability}
The code used to analyze the data and generate the figures in this paper is available at \url{https://github.com/onspatial/economic-sectors-mobility-data}
\vspace{-0.3cm}
\bibliographystyle{unsrt}
\bibliography{main}

\begin{thebibliography}{10}

\bibitem{richardson1969regional}
H.~W. Richardson.
\newblock {\em Regional economics. Location theory, urban structure and regional change.}
\newblock London: Weidenfeld and Nicolson., 1969.

\bibitem{delamater2016spatiotemporal}
Paul~L Delamater, Timothy~F Leslie, and Y~Tony Yang.
\newblock A spatiotemporal analysis of non-medical exemptions from vaccination: California schools before and after sb277.
\newblock {\em Social science \& medicine}, 168:230--238, 2016.

\bibitem{redding2017quantitative}
Stephen~J Redding and Esteban Rossi-Hansberg.
\newblock Quantitative spatial economics.
\newblock {\em Annual Review of Economics}, 9(1):21--58, 2017.

\bibitem{kim2019simulating}
Joon-Seok Kim, Hamdi Kavak, Umar Manzoor, Andrew Crooks, Dieter Pfoser, Carola Wenk, and Andreas Z{\"u}fle.
\newblock Simulating urban patterns of life: A geo-social data generation framework.
\newblock In {\em Proceedings of the 27th ACM SIGSPATIAL international conference on advances in geographic information systems}, pages 576--579, 2019.

\bibitem{hurley2024spatiotemporal}
Brendan~J Hurley and Timothy~F Leslie.
\newblock Spatiotemporal variograms as neighborhood definers.
\newblock {\em Geographical Analysis}, 56(2):404--424, 2024.

\bibitem{amiri2023massive}
Hossein Amiri, Shiyang Ruan, Joon-Seok Kim, Hyunjee Jin, Hamdi Kavak, Andrew Crooks, Dieter Pfoser, Carola Wenk, and Andreas Zufle.
\newblock Massive trajectory data based on patterns of life.
\newblock In {\em Proceedings of the 31st ACM International Conference on Advances in Geographic Information Systems}, pages 1--4, 2023.

\bibitem{amiri2024geolife+}
Hossein Amiri, Richard Yang, and Andreas Z{\"u}fle.
\newblock Geolife+: Large-scale simulated trajectory datasets calibrated to the geolife dataset.
\newblock In {\em Proceedings of the 7th ACM SIGSPATIAL International Workshop on GeoSpatial Simulation}, pages 25--28, 2024.

\bibitem{amiri2024urban}
Hossein Amiri, Ruochen Kong, and Andreas Z{\"u}fle.
\newblock Urban anomalies: A simulated human mobility dataset with injected anomalies.
\newblock In {\em Proceedings of the 1st ACM SIGSPATIAL International Workshop on Geospatial Anomaly Detection}, pages 1--11, 2024.

\bibitem{pesavento2020data}
John Pesavento, Andy Chen, Rayan Yu, Joon-Seok Kim, Hamdi Kavak, Taylor Anderson, and Andreas Z{\"u}fle.
\newblock Data-driven mobility models for covid-19 simulation.
\newblock In {\em Proceedings of the 3rd ACM SIGSPATIAL International Workshop on Advances in Resilient and Intelligent Cities}, pages 29--38, 2020.

\bibitem{kupfer2021using}
John~A Kupfer, Zhenlong Li, Huan Ning, and Xiao Huang.
\newblock Using mobile device data to track the effects of the covid-19 pandemic on spatiotemporal patterns of national park visitation.
\newblock {\em Sustainability}, 13(16):9366, 2021.

\bibitem{safegraph}
SafeGraph.
\newblock https://www.safegraph.com, 2025.
\newblock Accessed: [2025-05-05].

\bibitem{advanresearch}
AdvanResearch.
\newblock Foot traffic / weekly patterns [dataset], 2025.
\newblock \url{https://doi.org/10.82551/X1PP-1F65}.

\bibitem{deweydata}
DeweyData.
\newblock https://www.deweydata.io/, 2025.
\newblock Accessed: [2025-05-05].

\bibitem{chang2021mobility}
Serina Chang, Emma Pierson, Pang~Wei Koh, Jaline Gerardin, Beth Redbird, David Grusky, and Jure Leskovec.
\newblock Mobility network models of covid-19 explain inequities and inform reopening.
\newblock {\em Nature}, 589(7840):82--87, 2021.

\bibitem{elarde2021change}
Justin Elarde, Joon-Seok Kim, Hamdi Kavak, Andreas Z{\"u}fle, and Taylor Anderson.
\newblock Change of human mobility during covid-19: A united states case study.
\newblock {\em PloS one}, 16(11):e0259031, 2021.

\bibitem{kang2020multiscale}
Yuhao Kang, Song Gao, Yunlei Liang, Mingxiao Li, Jinmeng Rao, and Jake Kruse.
\newblock Multiscale dynamic human mobility flow dataset in the us during the covid-19 epidemic.
\newblock {\em Scientific data}, 7(1):390, 2020.

\bibitem{griffin2020mitigating}
Greg~P Griffin, Megan Mulhall, Chris Simek, and William~W Riggs.
\newblock Mitigating bias in big data for transportation.
\newblock {\em Journal of Big Data Analytics in Transportation}, 2(1):49--59, 2020.

\bibitem{li2024understanding}
Zhenlong Li, Huan Ning, Fengrui Jing, and M~Naser Lessani.
\newblock Understanding the bias of mobile location data across spatial scales and over time: a comprehensive analysis of safegraph data in the united states.
\newblock {\em Plos one}, 19(1):e0294430, 2024.

\bibitem{osf}
Timothy Leslie, Hossein Amiri, and Andreas Z\"ufle.
\newblock Exploring economic sectoral dynamics through high-resolution mobility data, 2025.

\bibitem{nps}
{National Park Service}.
\newblock https://www.nps.gov, 2025.
\newblock Accessed: [2025-02-10].

\end{thebibliography}
\end{document}